\input{aipcheck}

\documentclass[
    ,final            
  ]
  {aipproc}

\layoutstyle{6x9}

\begin{document}

\title{Quark-Hadron Duality in Neutron Spin-Structure\\
 and\\
$g_2$ moments at intermediate $Q^2$}


\author{P. Solvignon}{
  address={Physics Division, Argonne National Laboratory, Argonne, IL 60439}
}

\begin{abstract}
Jefferson Lab experiment E01-012 measured the $^3$He spin-structure functions and 
virtual photon asymmetries in the resonance region in the momentum transfer range 
1.0$<$Q$^2$$<$4.0 (GeV/c)$^2$. Our data, when compared with existing deep inelastic 
scattering data, were used to test quark-hadron duality in $g_1$ and A$_1$ for 
$^3$He and the neutron. In addition, preliminary results on the $^3$He spin-structure 
function $g_2$, on the Burkhardt-Cottingham sum rule and on higher twist effects 
through the $x^2$-weighted moment $d_2$ of the neutron were presented. 

\end{abstract}

\maketitle



\section{Quark-hadron duality}
In 1970, Bloom and Gilman~\cite{Bloom:1970xb} observed that structure function 
data taken at the Stanford Linear Accelerator Center (SLAC) in the resonance 
region average to the scaling curve of deep inelastic scattering (DIS). From the 
time the observation of quark-hadron duality was made, substantial efforts were 
put into a theoretical explanation for this phenomenon. In 
addition, the idea of a dual behavior between quarks and hadrons was extended to 
spin structure function $g_1$. 
Recent data from Jefferson Lab (JLab)~\cite{Niculescu:2000tj,Bosted:2006gp,Dharmawardane:2006zd,Wesselmann:2006mw} and DESY~\cite{Airapetian:2002rw} on the proton in the resonance 
region indicate the onset of duality at momentum transfers ($Q^2$) as low as 0.5 and 
1.6 (GeV/c)$^2$ for the unpolarized and polarized structure functions, respectively.

Carlson and Mukhopadhyay~\cite{Carlson:1998gf} showed within perturbative QCD 
that, at large $Q^2$ and as $x$ goes to 1, structure functions in the resonance 
region behaves the same way as in DIS region\footnote{In this proceeding, we call 
$x$ the Bj$\ddot{\rm o}$rken variable which is defined, in the parton model, as the 
nucleon momentum fraction carried by the struck parton.}. In the high $x$ region, 
the photon is more likely to interact with the quark having the same helicity as the 
nucleon. This implies that both $g_1$ and the unpolarized structure function $F_1$ 
behave as $(1-x)^3$ when $x \rightarrow 1$. 
The virtual photon-nucleon asymmetry $A_1$ is expected~\cite{Farrar:1975yb} to tend 
to 1 as $x \rightarrow 1$ in the scaling region. Carlson and Mukhopadhyay, considering 
resonant contributions and non-resonant background, predict the same behavior in the 
resonance region at large enough momentum transfer. 
Recently, Close and Melnitchouk~\cite{Close:2003wz} studied three different 
conditions of SU(6) symmetry breaking in the resonance region under which 
predictions of the structure functions at large $x$ lead to the same behavior as 
in the DIS region. (See Ref.~\cite{Melnitchouk:2005zr} for a detailed review of 
quark-hadron duality). 

Because of their different resonance spectra, it is expected, in certain theoretical 
models, that the onset of duality for the neutron will happen at lower momemtum 
transfer than for the proton. Now that precise neutron spin-structure 
data~\cite{ZHENG} in the DIS region are available at large $x$, data in the 
resonance region are needed in order to test quark-hadron duality on the neutron 
spin-structure function $g_1$. The goal of experiment E01-012 was to provide such data 
on the neutron ($^3$He) in the moderate $Q^2$ region up to $Q^2$ = 4.0 (GeV/c)$^2$ 
where duality is expected to hold.

In 2003, experiment E01-012 took data in Hall A at JLab. It was an 
inclusive measurement of longitudinally polarized electrons scattering off a 
longitudinally or transversely polarized $^3$He target~\cite{Alcorn:2004sb}. Asymmetries 
and cross section differences were measured in order to extract the spin-structure 
function $g_1$:
\begin{equation}
g_1(E,E^{\prime},\theta) = \frac{M Q^2 \nu}{4 \alpha_e^2} \frac{E}{E^{\prime}} 
\frac{1}{E+E^{\prime}} \bigg[\Delta \sigma_{\parallel}(E,E^{\prime},\theta) + 
\tan\frac{\theta}{2}~\Delta \sigma_{\perp}(E,E^{\prime},\theta) \bigg]
\label{eq:g1}
\end{equation}
where the superscript $\parallel$ ($\perp$) represents the configuration between the 
incident electron longitudinal spin direction and the longitudinal (transverse) target 
spin direction. The quantities $E$, $E^{\prime}$ and $\theta$ correspond to the incident 
and scattered electron energies and the scattering angle, respectively. Also in 
Eq.~\ref{eq:g1}, $M$ is the mass of the target, $\nu$ is the energy transfer to the 
target, $\alpha_e$ is the fine structure constant.
Note that our data allows a direct extraction of $g_1$ (and $g_2$, see Eq.~\ref{eq:g2}) 
without the need of an external input. All details on the experimental setup and the 
analysis steps can be found in~\cite{Solvignon:2006qc}.

\begin{figure}[t]
  \includegraphics[height=.6\textheight]{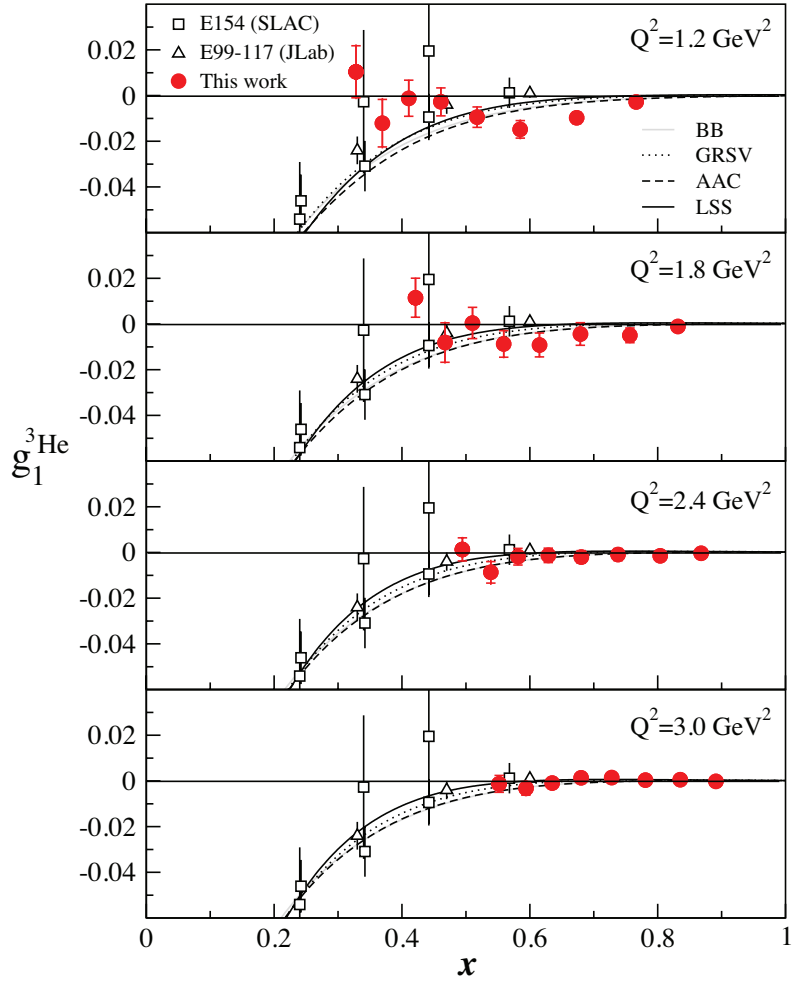}
  \label{fig_g1}
  \caption{The spin-structure function $g_1^{\rm ^3He}$ in the resonance
region at four $Q^2$-values. The error bars represent the total uncertainties with the
inner part being statistical only. Also plotted are the DIS world data from experiments
E154 at SLAC~\cite{Abe:1997cx} and E99-117 at JLab~\cite{ZHENG} which are at different 
$Q^2$ than our resonance data. The curves were generated from the NLO parton distribution
functions of Ref.~\cite{Blumlein:2002be,Gluck:2000dy,Goto:1999by,Leader:2005kw} to which 
target-mass correction were applied}
\end{figure}

The structure functions $g_1$ and $g_2$ were generated for the three incident energies and
two scattering angles, and then, were interpolated to constant $Q^2$. In Fig.~\ref{fig_g1}, 
the results from E01-012 on the spin-dependent structure function $g_1^{\rm ^3He}(x,Q^2)$ 
(per nucleon\footnote{In Eq.~\ref{eq:g1}, the proton mass was used instead of the mass of
the $^3$He nucleus.} are shown compared to parametrizations of parton distribution functions 
from four different groups~\cite{Blumlein:2002be,Gluck:2000dy,Goto:1999by,Leader:2005kw}, 
taken at Next-to-Leading Order (NLO). 
Target-mass corrections were applied to the DIS parametrization following the prescription 
of Ref.~\cite{Sidorov:2006fi}. These plots indicate that our resonance region data approach
the DIS parametrizations with increasing $Q^2$. 

Note that the DIS parametrizations of $g_1^{\rm ^3He}$ were generated using the proton 
and neutron $g_1$ parametrizations and the effective polarization equation~\cite{Bissey:2001cw}:
\begin{equation} 
g_1^{\rm ^3He} = P_n~g_1^n + 2 P_p~g_1^p
\label{eq:effpol}
\end{equation}
where $P_n = 0.86 \pm 0.02$ and $P_p = -0.028 \pm 0.004$ are the effective polarizations 
of the neutron and the proton in $^3$He, respectively~\cite{Friar:1990vx}.

In order to quantitatively study quark-hadron duality in the spin-structure function $g_1$, 
a partial integration is performed:
\begin{equation} 
\tilde{\Gamma}_1(Q^2) = \int_{x_{min}}^{x_{max}} dx~g_1(x,Q^2)
\label{eq:gam1}
\end{equation}
The partial moment for the neutron was extracted from the partial moment of $^3$He 
using Eq.~\ref{eq:effpol} by replacing the $g_1$-quantities by their partial moments 
$\tilde{\Gamma}_1$. This procedure was shown to be valid in 
Ref.~\cite{CiofidegliAtti:1996cg}. {\it Global} duality is defined as the partial
moment over the entire resonance region, from pion threshold (with missing mass $W = 
1.079$ GeV corresponding to $x_{max}$) to $W = 2.0$ GeV (corresponding to $x_{min}$). 
As for {\it local} duality, the partial integral is taken over a set of resonances. 

\begin{figure}[t]
  \includegraphics[height=.55\textheight]{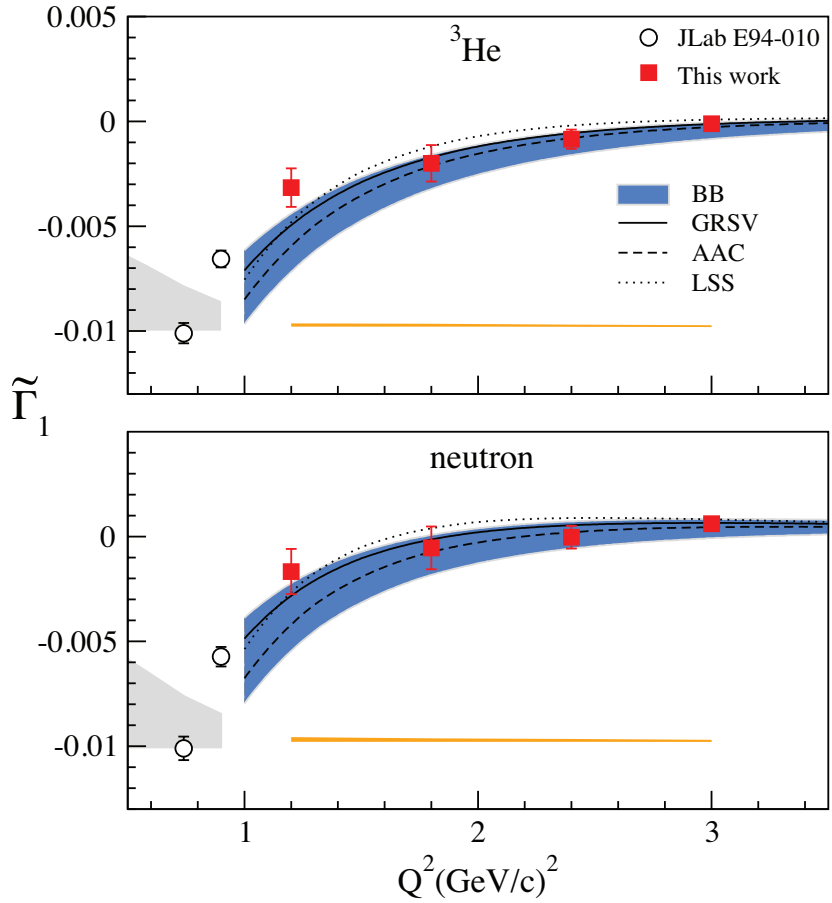}
  \label{fig_gam1}
  \caption{The partial $g_1$ first moment $\tilde{\Gamma}_1^{\rm ^3He}$ and 
$\tilde{\Gamma}_1^n$: test of spin duality on ${\rm ^3He}$ (top) and neutron (bottom). The
recent data from E01-012 are plotted with red squares, with the error bars being statistical 
only and the orange band being the absolute systematic uncertainty.
Also plotted are the DIS parameterizations of Bl\"{u}mlein and 
B\"{o}ttcher~\cite{Blumlein:2002be} (blue band), GRSV~\cite{Gluck:2000dy} (solid curve), 
AAC~\cite{Goto:1999by} (dashed curve) and LSS~\cite{Leader:2005kw} (dotted curve) after 
applying target-mass corrections. The open circles are data from JLab E94-010~\cite{E94010}
with the absolute systematic uncertainty represented by the grey band.}
\end{figure}

For all $Q^2$ settings of E01-012, the data cover a $x$-range corresponding to a $W$-range 
extending from the pion threshold to $W = 1.095$ GeV. Therefore we performed the integration
of Eq.~\ref{eq:gam1} over this $x$-range for our resonance data and for the DIS 
parametrizations shown in Fig.~\ref{fig_g1}. The result of this quantitative test of 
quark-hadron duality is shown in Fig.~\ref{fig_gam1}. We can see a clear confirmation
that global quark-hadron duality holds at least down to $Q^2 = 1.8$ (GeV/c)$^2$ for 
$^3$He and the neutron. Note that global duality was experimentally observed for 
the proton and the deuteron spin structure 
functions~\cite{Bosted:2006gp,Dharmawardane:2006zd,Wesselmann:2006mw} for $Q^2$ above 
1.7 (GeV/c)$^2$.

\vspace{0.4cm}
We also studied quark-hadron duality on the virtual photon-nucleon asymmetry $A_1$, which 
can be expressed from the parallel and perpendicular asymmetries ($A_{\parallel}$ and 
$A_{\perp}$) as follows:
\begin{equation} 
A_1 = \frac{A_{\parallel}}{D(1+\eta \xi)} - \frac{\eta A_{\perp}}{d(1+\eta\xi)}
\label{eq:a1}
\end{equation}
The variables $\eta$ and $\xi$ depend on the kinematics, and $D$ and $d$ are functions of 
the longitudinal to transverse cross section ratio $R(x,Q^2)$. Details on our evaluation
of $R$ for $^3$He can be found in~\cite{Solvignon:2008hk}.
\begin{figure}[t]
  \includegraphics[height=.48\textheight]{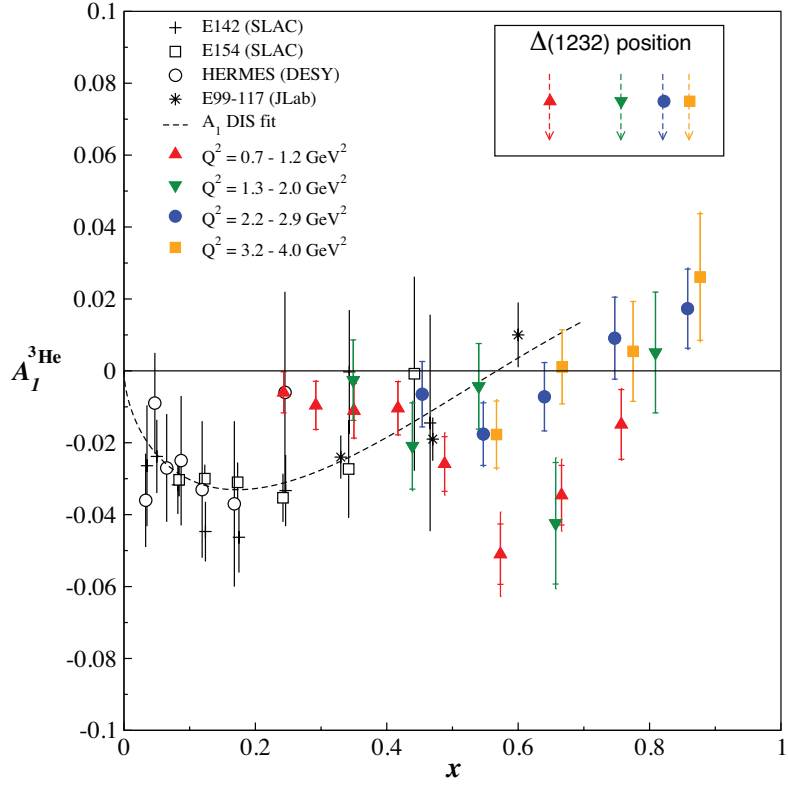}
  \label{fig_a1}
  \caption{The virtual photon-nucleon asymmetry $A_1^{\rm ^3He}$ in the resonance region. 
DIS data are from SLAC E142~\cite{Anthony:1996mw}, E154~\cite{Abe:1997cx}, from DESY 
experiment HERMES~\cite{Ackerstaff:1999ey} and from JLab E99-117~\cite{ZHENG}. The error 
bars represent the total uncertainties with the inner part being statistical only. The 
curve represents a fit to the $A_1^{\rm ^3He}$ DIS data. The arrows in the black frame 
point to the $\Delta$(1232) peak position for each of our data sets}
\end{figure}

The virtual photon-nucleon asymmetry $A_1^{\rm{^3He}}$ was extracted in the resonance 
region from our data at four different $Q^2$-ranges and is shown in Fig.~\ref{fig_a1}. 
The position of the $\Delta$(1232) resonance is indicated for each subset of data. 
The most noticeable feature is the negative contribution of the $\Delta$(1232) resonance 
at low $Q^2$. It has been argued~\cite{Carlson:1998gf,Close:2003wz} that quark-hadron 
duality should not work in the $\Delta$-region at low $Q^2$. However, at $Q^2$ above 
2.0 (GeV/c)$^2$, the dominant negative bump at the location of $\Delta$(1232) seems 
to vanish. Furthermore the results from these higher $Q^2$ settings show that the trend
of $A_1^{\rm{^3He}}$ goes to positive values with increasing $x$, as previously reported 
from the DIS world data. Our $A_1^{\rm{^3He}}$ results from the two highest $Q^2$ 
settings agree well with each other showing no strong $Q^2$-dependence.

The polarized $^3$He target was used in this experiment as an effective neutron 
target. Because of the dominant S-state of $^3$He where the two protons have their 
spins anti-aligned, we can expect neutron spin-structure functions to show similar 
behavior as observed for $^3$He structure functions here. Work is ongoing to extract 
the neutron $A_1$ results from the $^3$He results using the new convolution approach 
of~\cite{Kulagin:2008fm,Kahn:2008nq}. 

\section{The other spin-structure function}

In the naive parton model, the spin-structure function $g_2$ does not exist.
However the QCD parton model predicts a non-zero value for $g_2$. In the Operator 
Product Expansion (OPE) framework, both twist-two and higher twists operators contribute 
to $g_2$ as follows:

\begin{equation}
g_2(x,Q^2) = g_2^{WW}(x,Q^2) + \bar{g_2}(x,Q^2)
\label{eq:g2all}
\end{equation}

where $\bar{g_2}$ is the twist-three (and higher) contribution. The twist-two part of 
$g_2$ can be expressed using the Wandzura-Wilczek formula defined entirely from the 
knowledge of the spin-structure function $g_1$:

\begin{equation}
g_2^{WW}(x,Q^2) = - g_1(x,Q^2) + \int_x^1 \frac{dy}{y} g_1(y,Q^2)
\label{eq:g2ww}
\end{equation}

\begin{figure}[t]
  \includegraphics[height=.55\textheight]{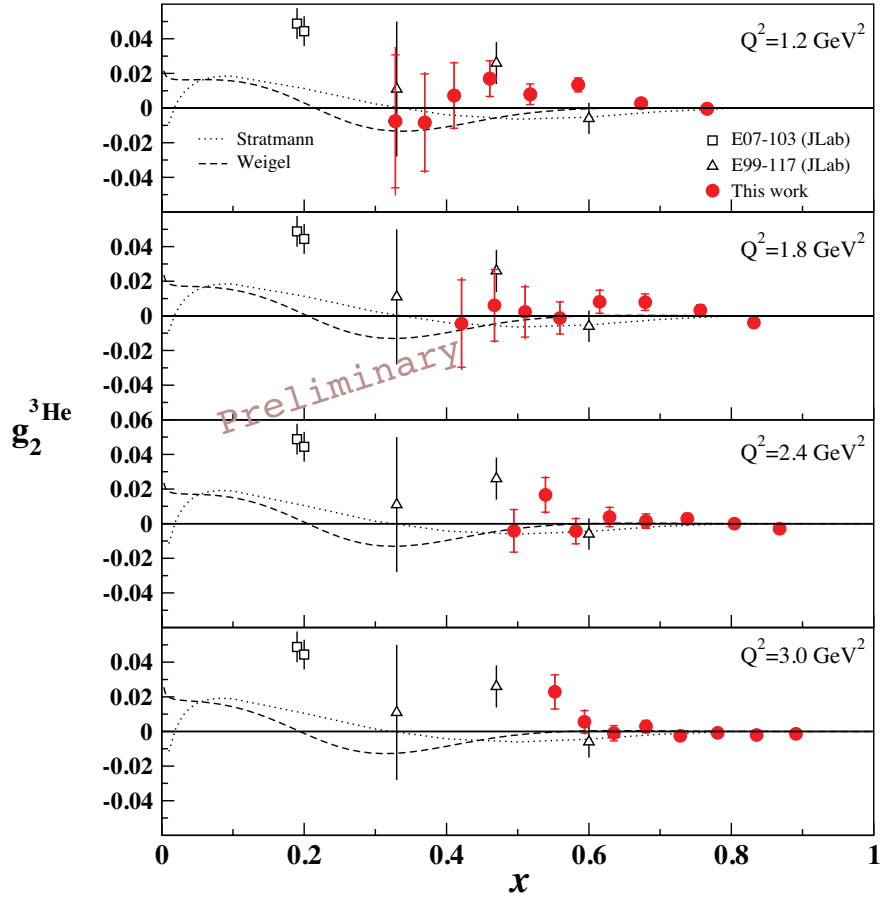}
  \label{fig_g2}
  \caption{The spin-structure function $g_2^{\rm ^3He}$ (per nucleon) in the resonance
region at four $Q^2$-values. The error bars represent the total uncertainties with the
inner part being statistical only. Also plotted are the DIS world data from JLab 
experiments E97-103~\cite{Kramer:2005qe} and E99-117 at JLab~\cite{ZHENG} which are at
different $Q^2$ than our resonance data. The dashed and dotted curves are calculations 
from the chiral soliton model~\cite{Weigel:1996jh} and from the bag 
model~\cite{Stratmann:1993aw} respectively.}
\end{figure}

The twist-three part of $g_2$ is not $1/Q$ suppressed compared to the $g_2^{WW}$ 
(twist-two) part. Therefore $g_2$ presents the unique advantage of the possible 
direct extraction of the twist-three contribution to the nucleon structure at high
$Q^2$, where higher than twist-three contributions are suppressed. 

Experimentally, one can perfom a model-independent measurement of $g_2$ by scattering
longitudinally polarized electron beam on a target with both longitudinal and transverse 
polarizations. The extraction of $g_2$ from the polarized cross section differences is
done following this formula:

\begin{equation}
g_2 = \frac{M Q^2 \nu^2}{4 \alpha_e^2} \frac{1}{2 E^{\prime}} \frac{1}{E+E^{\prime}} 
\bigg[- \Delta \sigma_{\parallel} + \frac{E+E^{\prime} \cos \theta}{E^{\prime} 
\sin \theta}~\Delta \sigma_{\perp} \bigg]
\label{eq:g2}
\end{equation}

Figure~\ref{fig_g2} presents the preliminary results on $g_2^{\rm ^3He}$ from E01-012 
at four $Q^2$ values. Also plotted are calculations from chiral soliton 
model~\cite{Weigel:1996jh} and from the bag model~\cite{Stratmann:1993aw} for $g_2^{\rm ^3He}$ 
in the DIS region. In the $x$-range covered by our data, we can see that $g_2^{\rm ^3He}$ 
is small and in agreement with the two theoretical models. 
\subsection{The ${\bf x^2}$-weighted moment ${\bf d_2}$}

In the OPE framework~\cite{Wilson:1969zs,Kodaira:1979ib}, information on the quark and 
gluon fields are contained in operators which can be {\it twist}-expanded in terms of 
$1/Q^{\tau}$. The twist $\tau$ is defined as the mass dimension minus the spin of the 
operator. From here, several sum rules can be generated from the spin-structure functions 
$g_1$ and $g_2$:

\begin{equation}
\int_0^1 dx~x^n g_1(x,Q^2) = \frac{1}{2} a_n~~~~~~n = 0,2,4,...
\label{eq:twist1}
\end{equation}
\begin{equation}
\int_0^1 dx~x^n g_2(x,Q^2) = \frac{1}{2} \frac{n}{n+1} (d_n - a_n) ~~~~~~n = 2,4,...
\label{eq:twist2}
\end{equation}

with $a_n$ ($d_n$) are the twist-two (higher twists) reduced matrix elements. From 
Eqs.~\ref{eq:twist1}~and~\ref{eq:twist2}, we can extract the twist-three (and higher) 
matrix element $d_2$: 
\begin{equation}
d_2(Q^2) = \int_0^1 dx^2 \Big[2 g_1(x,Q^2) + 3 g_2(x,Q^2)\Big]= 3 \int_0^1 dx^2 
\bar{g}_2(x,Q^2)
\label{eq:d2}
\end{equation}
The leading twist quantities can be easily compared to naive parton model predictions. 
Higher twist effects are due to quark-quark and quark-gluon interactions. The twist-three
quantities correspond to quark-gluon correlations.

\begin{figure}[t]
  \includegraphics[height=.4\textheight]{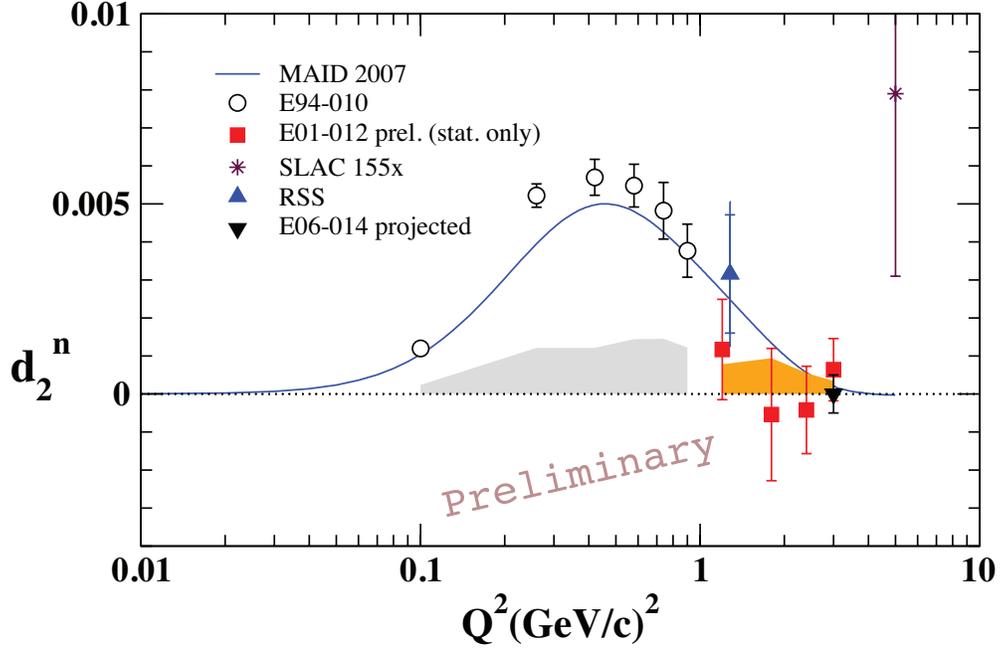}
  \label{fig_d2}
  \caption{Preliminary results on the {\it resonance contribution} to the neutron 
$x^2$-weighted moment $d_2$ from E01-012. The error bars are statistical only and the 
band represents the experimental systematics. Data from JLab experiments 
E94-010~\cite{E94010} and RSS~\cite{Slifer:2008xu} are shown. For comparison to the 
resonance contribution, we plotted the MAID model~\cite{Drechsel:2007if}. Also plotted 
are the {\it total} $d_2$ from SLAC experiment E155x~\cite{Anthony:2002hy} and the 
projected result from JLab E06-014~\cite{E06014}, currently under analysis.}
\end{figure}

Fig.~\ref{fig_d2} shows the preliminary results for resonance region contribution to 
$d_2^n$  from E01-012, and also from earlier JLab experiments E94-010~\cite{E94010}, 
RSS~\cite{Slifer:2008xu}. It is found to be very small for $Q^2$ above 1 (GeV/c)$^2$. 


Prediction from lattice QCD calculation~\cite{Gockeler:2000ja} has for the neutron 
$d_2 = -0.001 \pm 0.003$ at $Q^2 = 5$ and $10$ (GeV/c)$^2$ with a $Q^2$-evolution close to 
constant down to $Q^2 = 2$ (GeV/c)$^2$. This could mean that the unmeasured part of $d_2$ 
from E01-012 at $Q^2 = 3$ (GeV/c)$^2$ would be also very small. JLab experiment 
E06-014~\cite{E06014} should be able to tell us the answer in the next couple of years.

Also, it is really exciting to see the good agreement between E01-012 and RSS data since 
they come from two different experimental setups and two different targets: polarized $^3$He 
for E01-012 and polarized $^2$H for RSS.
\subsection{The Burkhardt-Cottingham Sum Rule}

The Burkhardt-Cottingham (BC) sum rule~\cite{Burkhardt:1970ti} is a super-convergence 
relation derived from dispersion relation in which the virtual Compton helicity amplitude 
$S_2$ falls off to zero more rapidly than $\frac{1}{\nu}$ as $\nu \rightarrow \infty$.
The sum rule is expressed as follows: 
\begin{equation}
\Gamma_2(Q^2) \equiv \int_0^1 dx~g_2(x,Q^2) = 0,
\label{eq:bc}
\end{equation}
and is predicted to be valid at all $Q^2$.
The validity of the sum rule derived through assumptions of Regge theory has been 
questionned~\cite{Jaffe:1989xx}. Also it can be seen from Eq.~\ref{eq:twist2} that the 
BC sum rule cannot be extracted from the OPE due to the non-existent $n=0$ expansion
of $g_2$-moments.

\begin{figure}[t]
  \includegraphics[height=.4\textheight]{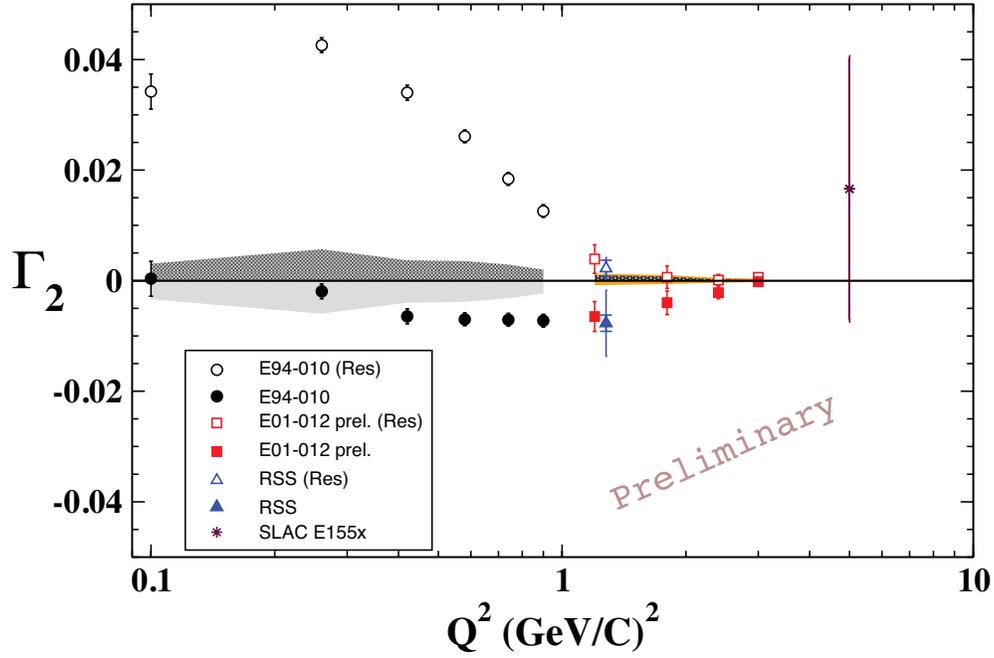}
  \label{fig_bc}
  \caption{Preliminary results on the Burkhardt-Cottingham sum rule on the neutron from
E01-012 (filled squares). The error bars are statistical only, the upper band represents 
the experimental systematics and the lower band the uncertainties on the unmeasured part
of the sum rule. The open square data are the measured part of the integral as was perfomed
by experiment E01-012. Also plotted are data from JLab experiments E94-010~\cite{E94010} 
and RSS~\cite{Slifer:2008xu}, with also the measured part of the integral represented by 
open symbols and the sum rule with filled symbols, and SLAC experiment 
E155x~\cite{Anthony:2002hy}.} 
\end{figure}

Preliminary data from E01-012 on the BC sum rule are shown in Fig.~\ref{fig_bc}. Also
shown are data from JLab experiments E94-010~\cite{E94010} and RSS~\cite{Slifer:2008xu}.
All these experiments were concentrated on the resonance region and therefore have 
measured only the resonance part of Eq.~\ref{eq:bc}. In order to generate the full 
integral, the unmeasured elastic and DIS contributions need to be added. For the elastic 
part, we used the parametrization from Ref.~\cite{Mergell:1995bf}. However, for the DIS
contribution, we used $g_2^{WW}$ which can be evaluated from our own $g_1$ data. A 
conservative systematic uncertainty was associated with this approximation and more 
systematic studies are underway looking at using different theoretical models to evaluate
the low $x$ unmeasured part of the integral.

Nonetheless, at this point in our analysis, we can see in Fig.~\ref{fig_bc} data approaching
the BC sum rule with increasing $Q^2$.
We can also see the good agreement between E01-012 and RSS data.

\section{Conclusion}
Experiment E01-012 provides spin-structure data in the resonance region for the neutron 
($^3$He) for $1.0 < Q^2 < 4.0$ (GeV/c)$^2$ and $0.30 < x < 0.85$. Quark-hadron duality
was found to hold globally for the neutron and $^3$He spin-structure function $g_1$ at 
least down to $Q^2 = 1.8$ (GeV/c)$^2$. At $x < 0.60$, where DIS $A_1^{\rm{^3He}}$ data 
are available, a qualitative {\it local} test of quark-hadron duality was performed. The 
results show that $A_1^{\rm{^3He}}$ in the resonance region follows a similar behavior as 
$A_1^{\rm{^3He}}$ measured in the DIS region. The confirmation of quark-hadron duality for 
the neutron structure functions is important for a better understanding of the mechanism 
of quark-gluon and quark-quark interactions. Combined with already existing proton resonance 
data, a study of spin and flavor dependence of duality can be performed. 

Preliminary results from E01-012 show small values for the neutron $x^2$-weighted moment 
$d_2$ above $Q^2 \approx 1$ (GeV/c)$^2$.
Also, our results on the Burkardt-Cottingham sum rule is in good agreement with the existing
world data showing that the sum rule is valid at the two-sigma level for $Q^2$ between 0.1 
and 5.0 (GeV/c)$^2$. 

Finally, more results are expected to come from E01-012 as the extraction of $A_1^n$ in
the resonance from our data on $^3$He, the extended GDH sum rule, the Bjorken sum rule, 
etc.

\begin{theacknowledgments}
This work was supported by U.S. Department of Energy, Office of Nuclear Physics, under 
contract numbers DE-AC02-06CH11357 and DE-AC05-84ER40150 Modification No.~M175.
\end{theacknowledgments}



\bibliographystyle{aipprocl} 

\bibliography{spindual_g2}

\end{document}